\RequirePackage{fix-cm}
\documentclass[natbib,smallextended]{svjour3}       
\smartqed  

\usepackage{amsmath,amssymb,amsfonts}
\usepackage{algorithmic}
\usepackage{graphicx}
\usepackage{textcomp}

\usepackage{algorithm}
\usepackage{algorithmic}

  \usepackage{rotating}
        \usepackage{multirow}
        \usepackage{lscape}

\usepackage{xcolor}

 \journalname{}
\begin{document}

\title{ Recommendation System based on Semantic Scholar Mining and Topic modeling: A behavioral analysis of researchers from six conferences
}


\author{Hamed Jelodar \and Yongli Wang \and Mahdi Rabbani \and Ru-xin Zhao \and Seyedvalyallah Ayobi \and Peng Hu \and Isma Masood
}


\institute{H. Jelodar \at
              School of Computer Science and Technology, Nanjing University of Science and Technology, Nanjing 210094, China \\
              Tel.: +8615298394479\\
              \email{Jelodar@njust.edu.cn}           
           \and
           Y. Wang \at
              School of Computer Science and Technology, Nanjing University of Science and Technology, Nanjing 210094, China \\
              \email{YongliWang@njust.edu.cn}           
}

\date{Received: date / Accepted: date}

\maketitle

\begin{abstract}
Recommendation systems have an important place to help online users in the internet society. Recommendation Systems in computer science are of very practical use these days in various aspects of the Internet portals, such as social networks, and library websites. There are several approaches to implement recommendation systems, Latent Dirichlet Allocation (LDA) is one the popular techniques in Topic Modeling. Recently, researchers have proposed many approaches based on Recommendation Systems and LDA. According to importance of the subject, in this paper we discover the trends of the topics and find relationship between LDA topics and Scholar-Context-documents. In fact, We apply probabilistic topic modeling based on Gibbs sampling algorithms for a semantic mining from six conference publications in computer science from DBLP dataset. According to our experimental results, our semantic framework can be effective to help organizations to better organize these conferences and cover future research topics.

\end{abstract}

\section{Introduction}
\label{sec1}

Recommendation Systems in computer science are of very practical use these days in various aspects of the Internet Society. Generally, a recommendation system offers a new item to a user after specifying which of the users item is most similar to the new item. Recommendation systems have been widely used in many fields, the most typical one is E-commerce, which has a good development and application prospect. A successful recommendation system can significantly improve the revenue and sales of e-commerce companies, so as to promote their rapid development. For instance, Amazon shopping that Amazon recommends items to a user based on items the user previously visited, or YouTube Website that YouTube recommends movies to a user based on the users prior ratings and watching habits. There are several methods based one recommendation systems, such as content-based filtering, collaborative filtering, and hybridizations. Recently, researchers proposed various methods based on probabilistic topic modeling methods. Latent Dirichlet Allocation (LDA) is becoming a standard tool in topic modeling [1]. LDA is a generative probabilistic model broadly used in the information retrieval field. Recently, researchers have used topic modeling approaches based on LDA to build recommendation systems in various subjects, such as scientific paper recommendation [2-9], music and video Recommendation [10-19], location recommendation[20-26], hashtag recommendation [27-39], travel and tour recommedation[40-43], app recommendation [44-48], event recommendation[49-54], social networks and media [55-59].Forc example, from an applied perspective in the field of music and video recommendation, Yan and et al focused on the efficiency of user's information content on the online social network and provided a solution as a personalized video recommendation with considering users cross-network social and content data. They applied a topic model based on LDA for each user, that user as document and user's hashtags as word, with considering user information from Twitter. In this paper, we evaluate scholar publications from six conferences to discover hidden behavioral aspects and discover the trends of the topics and find relationship between LDA topics and paper features. In summary, this paper makes three main contributions:
\begin{itemize}
\item This study shows the application of the topic models for building a recommendation systems based on semantic scholar mining for predicting interesting research field from scholarly articles.
\item We consider six conference publications from DBLP Library.
\item We discover relationship between Scholar-Context-documents and topics, and also analyses the trends of the topics in various years
\end{itemize}

\section{Research related concepts} \label{topib}

Recommendation systems play increasingly significant role in online web services. Many recommendation systems rely on data mining; that is, attempting to discover useful patterns in large data sets. While such recommendation systems are helpful, it is not always practical to create, maintain, and use the large data sets they require. Recommendation systems are increasingly being used in various applications such as movie recommendation, mobile recommendation, article recommendation and etc. For instance, Amazon shopping that Amazon recommends items to a user based on items the user previously visited, or YouTube Website that YouTube recommend movies to a user based on the users previous ratings and watching habits. The aim of a recommendation system is to use historical data about the users behavior (e.g., their purchases as well as ratings on purchased items) and provide a list of items to each user such that they are likely to be purchased by the user in near future. The Content-Based (CB)[60] and Collaborative Filtering (CF) [61] are most important techniques in recommendation system.

\subsection{Semantic analysis and topic modeling in recommendation systems}
LDA developed has become a widely used topic modeling algorithm [1]. LDA assumes that each word in a document is produced in two steps. First, assuming that each document has it's own topic distribution, a topic is randomly drawn based on the document's topic distribution. Next, assuming that each topic has its own word distribution, a word is randomly drawn from the word distribution of the topic selected in the previous step. Repeating these two steps word by word generates a document. In recent years a considerable amount of research has addressed the task of defining models and systems for scientific papers recommendation; this trend has emerged as a natural consequence of the increasing growth of the number of scientific publications. For example, Youn and et al, proposed an approach to scientific articles'recommendation of user's interests based on a topic modeling framework. The authors, used a LDA model in order to extract the topics of the followees'tweets (followed Twitterers) and the paper titles [5]. They apply the Twitter-LDA algorithm simultaneously on the followees'tweets and the paper titles with the number of topics set to 200, they utilized the intersection of topics found in both paper titles and followees's tweets. Each followee of a user is ranked as follows:

\begin{equation} \label{GrindEQ__5_}
{Rank}_{f\textrm{f}\textrm{o}llowee}=\left(\sum_{t\in {Topics}_p}{\frac{n\left(t,T_f\right)}{\left|T_f\right|}}\right)*|{Topics}_p\bigcap {Topics}_f|
\end{equation}

where $T_f$denotes all tweets by a followee,

${Topics}_p,\ $denotes the set of topics defining the titles of scientific
articles,

${Topics}_f,\ $denotes the set of topics defining the tweets of a followee
and\textbf{ }$n\left(t,T_f\right)\ $the number of times a particular topic
\textbf{`t'} from within occurs among the tweets of a followee. Based on the
ranking scores of all followees of a particular user,and obtained top-k researchers
followed by a target user. For evaluation approach, considered DBLP database as a
large academic bibliographic network.\\

\begin{table}

\centering
\caption{Impressive works LDA-based on paper recommendation}
\label{tab:1}       

\resizebox{12cm}{!} {

\begin{tabular}{lllll}
\hline\noalign{\smallskip}
Study & Year & Purpose & Method & Dataset\\
\noalign{\smallskip}\hline\noalign{\smallskip}

[5] & 2014 & Present an approach to utilize this & LDA, Twitter-LDA  & DBLP dataset\\
         & & valuable information source to suggest\\
          & & scientific articles\\

[2]  & 2017 & A scientific paper recommendation approach &  LDA, Gibbs sampling   & ArnetMiner Dataset\\
     &   &   &      & DBLP dataset\\

[3]  & 2017  & A Citation recommendation & LDA                           & ACL Anthology \\
     &       & present a topic model     &  Maximum A Posteriori (MAP)    & Network ,\\
     &       & combing with author link  &                                & DBLP \\
     &       & community\\

[4]  & 	2013 & 	A personalized recommendation & LDA, EM algorithm & 	Digg articles\\
     &       &   system for Digg articles     & 	              &     (digg.com)\\

[7] & 	2011 & 	A scientific articles recommendation & LDA, EM algorithm & 	CiteULike Dataset\\
     &       &   to users based on both content and\\
     &       &   other users ratings\\

\noalign{\smallskip}\hline
\end{tabular}
}

\end{table}

Also, some researchers, introduced a combined model based on traditional collaborative filtering and topic modeling and designed a novel algorithm to scientific articles recommendation for users from an online community, called CTR model. They considered LDA to initialize the CTR model, Infact they combined the matrix factorization and the LDA model, and is shown their approach better than the recommendations based on matrix factorization. For evaluation and test, Used a large dataset from a bibliography sharing service (CiteULike) [7]. Table 1, shown some impressive work based on LDA for paper recommendation.\\

\section{Research framework and Semantic mining}\

In this section, we present a semantic mining framework based on topic modeling to build a recommendation system for discovering interesting research field of scholars from six conferences publication, showed this semantic framework in figure 1. And also, we discover temporal topic trends of conferences in various years by advanced content mining. Figure 2 shows the applicability of our topic modelling for recommending interesting research field of scholars to director conferences and also showed a example of word distribution in scholar's articles in figure 3. The framework presented for this semantic scholar mining is performed in three steps:

\begin{figure}
  \centering
    \includegraphics[width=130mm]{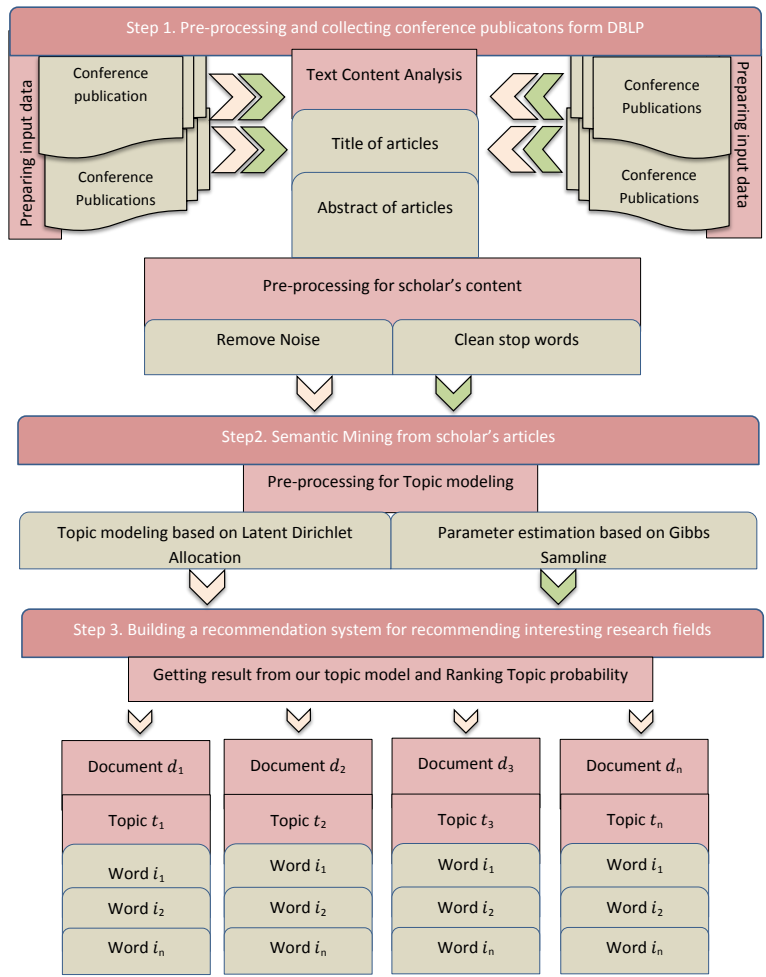}
  \caption{Research framework for recommending research field and semantic mining from scholar's articles.}
\end{figure}

\begin{figure}
  \centering
    \includegraphics[width=40mm]{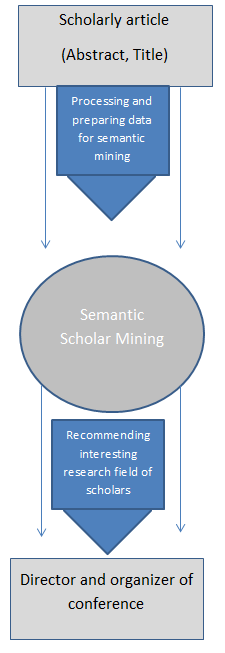}
  \caption{This figure shows the applicability of our topic model in a short process for conferences directors.}

\end{figure}

	Step 1.  We extracted six conference publications from DBLP website by only considering conferences for which data was available for years 2013-2017. DBLP is bibliographic information in computer science. After collecting scholar’s articles, the novel text information will be more concentrated. Then a stop list is used for removing stop-words from the set of terms in all documents. Stop-words are common words (such as “which”, “the”, “a” ,"as"), which do not effect on on our semantic analysis . These words do not support any beneficial data and eliminating them can decrease noise. Therefore, we need to find and remove them for avoiding interference with our framework. In this research, we use the stop list given by MALLET Topic Model to eliminate stop words by the step of word segmentation.\\

	Step 2.  After collecting articles and also preparing input data, a topic model based on Gibbs sampling applied for knowledge discovery and semantic mining in scholar’s articles. Topic modeling In our approach, the goal of topic modeling is to search the topic words related to a novel document so as to obtain the summary candidate sentences for building a recommendation system. LDA (Latent Dirichlet allocation) developed has become a widely used topic modeling [1]. LDA assumes that each word in a document is produced in two steps. First, assuming that each document has it's own topic distribution, a topic is randomly drawn based on the document's topic distribution. Next, assuming that each topic has its own word distribution, a word is randomly drawn from the word distribution of the topic selected in the previous step. Repeating these two steps word by word generates a document. Essentially, LDA reduces the extraordinary dimensionality of Scholar-Data from words to topics, based on word co-occurrences in the same document, similar to cluster analysis or principal component analysis. Therefore, we applied a LDA topic model with Gibbs sampling for semantic mining and discover relationship between Scholar-Context-documents and topics. According to LDA algorithm we define set of documents as Scholar-Context and words as topics. In mathematically, given the parameters $\alpha $ and $\beta $, the joint distribution of a topic mixture $\theta $, a set of \textit{N }topics z, anda set of \textit{N }words w is given by:\\

\begin{equation}
p(w|\alpha ,\beta )={\kern 1pt} \int  p(\theta |\alpha )\left(\prod _{n=1}^{N} \sum _{z_{n} } p(z_{n} |\theta )p(w_{n} |z_{n} ,\beta )\right)d\theta
\end{equation}

Finally, taking the generate of the marginal probability of every Scholar-articles in the corpus, the construction process probability of a corpus is determined as follows:
   (4)

\begin{equation} \label{GrindEQ__4_}
P(D|\alpha ,\beta )=\mathop{\mathop{\mathop{\prod  }\limits_{} }\limits^{M} }\limits_{m=1} {\kern 1pt} \mathop{\mathop{\mathop{\int  }\limits_{} }\limits^{} }\limits_{\theta _{m} } {\kern 1pt} \mathop{\mathop{\mathop{\int  }\limits_{} }\limits^{} }\limits_{\varphi _{k} } {\kern 1pt} p\left(\theta _{m} |\alpha \right)p(\varphi _{k} |\beta )\left(\mathop{\mathop{\mathop{\prod  }\limits_{} }\limits^{N_{m} } }\limits_{n=1\; } {\kern 1pt} \mathop{\mathop{\mathop{\sum  }\limits_{} }\limits^{} }\limits_{z_{m} } {\kern 1pt} p(z_{m,n} |\theta _{m} )p(w_{m,n} |\varphi _{z_{m,n} } )\right)d\theta _{m} d\varphi _{k} \; ,
\end{equation}

The parameters $\alpha $ and $\beta $ are corpus level  parameters, assumed to be sampled once in the process of generating a corpus. The variables $\theta _{m} $\textit{ }are document-level variables, sampled once per document. Finally, the variables \textit{zdn }and \textit{wdn }are word-level variables and are sampled once for each word in each document. Gibbs sampling is applied to learn the LDA model and each instance is then expressed with topic distributions. Then, we use a Gibbs sampler to allocate a new label(topic) to the word, by sampling:\\

\begin{equation} \label{GrindEQ__5_}
p(z_{i} =k|z_{i-1} ,d,w)\propto \frac{n_{wk} +\beta }{\mathop{\sum  }\limits_{v} n_{vk} +\beta _{v} } \left(n_{dk} +\alpha \right),
\end{equation}

where $n_{w,k} and\; n_{dk} $ are the respective counts of topics $k$ with words $w$ or in documents $d\; and\; \alpha \; and\; \beta $ hyperparameters as before. There are several methods to determine the model parameters that EM and Gibbs samplig are the most popular methods for parameter estimation. Gibbs sampling is a technique used to rapidly explore the space around a target distribution using repeated sampling.\\

	Step 3.  After semantic mining and extracting topics, in this paradigm, keywords generated based on ranking and high probability as a burst topic. In fact, Theses topic recommenced as research field of scholar data. These  result will be helpful for direct of conference  for understanding and scholar behavior analysis. And also we can discover trend topics in various years. Also, based on this framework and result obtained from this step, we answer two important question: \textit{Which research fields are very interesting for scholars in these conferences in computer science, between 2013-2017? What is relationship between LDA topics and articles features?}

\begin{figure}
  \centering
    \includegraphics[width=110mm]{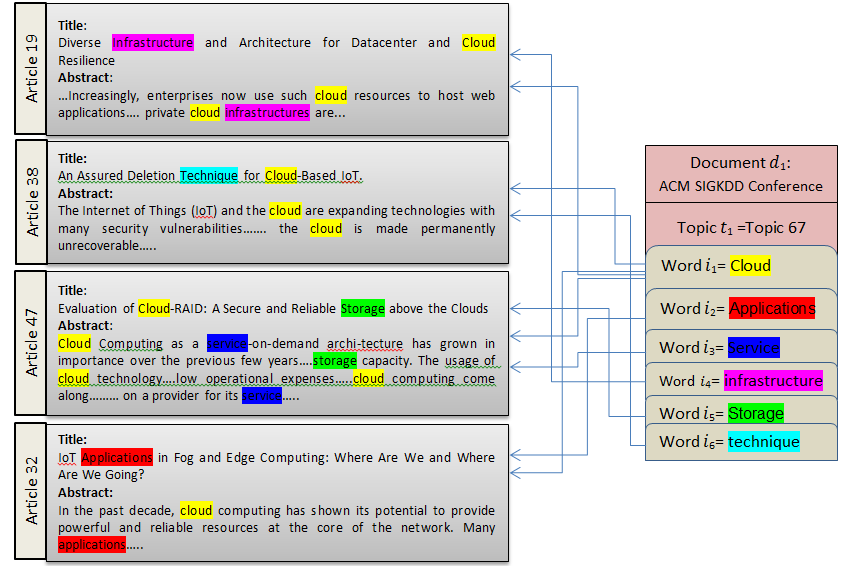}
  \caption{An example of word distribution in Scholar-document learned using the topic model. Some of the words in the document marked in different colors. }

\end{figure}

\section{Researcher behavior analysis from conferences publication}
We extracted six conference publications from DBLP website by only considering conferences for which data was available for years 2013-2017, including; AAAI, ACM SIGMOD, ACM SIGIR, ACM SIGKDD, ICCCN and ICSE Conferences. We pre-processed our data by a) removing stop-words, symbols, punctuations, and numbers b) down-casing the obtained words, and c)in the next step, prepared the dataset and then setting the parameters for topic modeling. It should be noted that in these experiments, we considered abstracts and titles from each article. In this paper, we used MALLET (http://mallet.cs.umass.edu/) to implement the inference and obtain the topic models. In addition, our full dataset is available at https://github.com/JeloH/Dataset\_DBLP.

\subsection{Parameter Settings}
 In this paper, all experiments were carried out on a machine running Windows 7 with CoreI3 and 4 GB memory. We learn a LDA model with 100 topics; $\alpha=0.01$, $\beta = 0.01$ and using Gibbs sampling as a parameter estimation. Related words for a topic are quite intuitive and comprehensive in the sense of  supplying a semantic short of a specific research field. The research topic associated with each topic for different years are quite representative. In addition, by doing analysis of our DBLP dataset, we found that all highly ranked research topic for various years have published papers in various conferences. For example, field of social media analysis (topic 62) for ACM SIGIR conference, we can see for Topic 62, published many papers in 2017 as a ranked first and in 2016 as a ranked second.

\subsection{Statistical analysis and Topic probability}
In this section, we provide the results and discovered topics of 100-topics for six conferences and we present temporal topic trends of conferences in various years.

\begin{table}[]
\centering
\caption{This result discovered topics of 100-topics from six conferences, each topic is shown with the top 20 words and identified 'research fields' that are covered at these conferences. Topics that have two concepts are in blue and red colors.
}
\label{my-label}

\resizebox{12cm}{!} {

\begin{tabular}{|l|l|l|l|c|l|c|l|}
\hline
\multicolumn{4}{|c|}{ACM SIGIR}                                                                                                                                                                                                                                                                                                                                                                                                                                                                                                        & \multicolumn{4}{c|}{ACM SIGKDD}                                                                                                                                                                                                                                                                                                                                                                                                                                                                                                                                                                                                                                             \\ \hline
\multicolumn{2}{|l|}{Information retrieval}                                                                                                                                                                                                         & \multicolumn{2}{l|}{Social media analysis}                                                                                                                                                                                                                                                    & \multicolumn{2}{c|}{{\color{red}Graph}, {\color{blue}Behavior  Analysis}}                                                                                                                                                                                                                                                                   & \multicolumn{2}{c|}{\begin{tabular}[c]{@{}c@{}}Decision systems\end{tabular}}                                                                                                                                                                                                                            \\ \hline
\multicolumn{2}{|c|}{Topic 59}                                                                                                                                                                                                              & \multicolumn{2}{c|}{Topic 62}                                                                                                                                                                                                                                                             & \multicolumn{2}{c|}{Topic 91}                                                                                                                                                                                                                                                                                      & \multicolumn{2}{c|}{Topic 99}                                                                                                                                                                                                                                                                                                                    \\ \hline

\begin{tabular}[c]{@{}l@{}}search\\query\\retrieval\\queries\\relevance\\documents\\social\\task\\study\\result\end{tabular} &
 \begin{tabular}[c]{@{}l@{}}improve\\important\\ir\\time\\number\\online\\algorithm\\list\\order\\address\end{tabular} &

 \begin{tabular}[c]{@{}l@{}}time\\large\\text\\present\\feedback\\state\\multiple\\twitter\\algorithms\\results\end{tabular} &
 \begin{tabular}[c]{@{}l@{}}measures\\general\\understanding\\standard\\design\\wikipedia\\called\\interface\\low\\engine\end{tabular} &

 \multicolumn{1}{r|}{\begin{tabular}[c]{@{}r@{}}Data\\{\color{red}model}\\large\\{\color{red}algorithm}\\{\color{red}network}\\real\\paper\\{\color{red}graph}\\scale\\experiments\end{tabular}} & \begin{tabular}[c]{@{}l@{}}{\color{blue}features}\\number\\{\color{blue}feature}\\process \\{\color{blue}classification}\\text\\{\color{blue}behavior}\\tasks\\level\\recent\end{tabular} &

  \multicolumn{1}{r|}{\begin{tabular}[c]{@{}r@{}}high\\multi\\art\\important\\learning\\systems\\techniques\\approaches\\scalable\\application\end{tabular}} & \begin{tabular}[c]{@{}l@{}}specific\\decision\\find\\finally\\big\\accurate\\dimensional\\specifically\\optimization\\low\end{tabular} \\ \hline

\multicolumn{2}{|l|}{Predicting information}                                                                                                                                                                                                         & \multicolumn{2}{l|}{Ad hoc information retrieval}                                                                                                                                                                                                                                                    & \multicolumn{2}{c|}{Pattern mining}                                                                                                                                                                                                                                                                   & \multicolumn{2}{c|}{Multiple nodes and Query Analysis}                                                                                                                                                                                                                            \\ \hline

\multicolumn{2}{|c|}{Topic 68}                                                                                                                                                                                                              & \multicolumn{2}{c|}{Topic 98}                                                                                                                                                                                                                                                             & \multicolumn{2}{c|}{Topic 29}                                                                                                                                                                                                                                                                                      & \multicolumn{2}{c|}{Topic 66}                                                                                                                                                                                                                                                                                                                    \\ \hline

\begin{tabular}[c]{@{}l@{}}show\\features\\demonstrate\\work\\behavior\\datasets\\propose\\methods\\dataset\\level\end{tabular} &
 \begin{tabular}[c]{@{}l@{}}classification\\prediction\\items\\outperforms\\recent\\participants\\scale\\generated\\studies\\explore\end{tabular} &

\begin{tabular}[c]{@{}l@{}}model\\proposed\\quality\\word\\results\\state\\attention\\graph\\trec\\interactions\end{tabular} &
 \begin{tabular}[c]{@{}l@{}}preferences\\mobile\\corpus\\networks\\Ad\\factorization\\embedding\\hashing\\specifically\\labels\end{tabular} &

 \multicolumn{1}{r|}{\begin{tabular}[c]{@{}r@{}}networks\\framework\\data\\prediction\\patterns\\datasets\\present\\work\\efficient \\search\end{tabular}} & \begin{tabular}[c]{@{}l@{}}key\\mining\\discovery\\efficiently\\make\\common\\location\\challenging\\node\\recommendation\end{tabular} &

  \multicolumn{1}{r|}{\begin{tabular}[c]{@{}r@{}}models\\proposed\\world\\analysis\\system\\multiple\\state\\existing\\online\\linear\end{tabular}} & \begin{tabular}[c]{@{}l@{}} \\web\\learn\\nodes\\outperforms\\evaluate\\structure\\interest\\global\\synthetic\\query\end{tabular} \\ \hline


\multicolumn{4}{|c|}{ACM SIGMOD}                                                                                                                                                                                                                                                                                                                                                                                                                                                                                                        & \multicolumn{4}{c|}{AAAI}                                                                                                                                                                                                                                                                                                                                                                                                                                                                                                                                                                                                                                             \\ \hline

\multicolumn{2}{|l|}{Management of Data}                                                                                                                                                                                                         & \multicolumn{2}{l|}{{\color{red}Aggregation operators}, {\color{blue}Storage management}}                                                                                                                                                                                                                                                   & \multicolumn{2}{c|}{NLP and Knowledge Representation}                                                                                                                                                                                                                                                                   & \multicolumn{2}{c|}{\begin{tabular}[c]{@{}c@{}}{\color{blue}Large scale models }, {\color{red}Unsupervised Learning}\end{tabular}}

\\ \hline
\multicolumn{2}{|c|}{Topic 11}                                                                                                                                                                                                              & \multicolumn{2}{c|}{Topic 45}                                                                                                                                                                                                                                                             & \multicolumn{2}{c|}{Topic 23}                                                                                                                                                                                                                                                                                      & \multicolumn{2}{c|}{Topic 37}

                                                                                                                                                                                                                                                                                                                  \\ \hline

\begin{tabular}[c]{@{}l@{}}data\\query\\based\\queries\\processing\\graph\\memory\\paper\\real\\systems\\efficient\end{tabular}
&
 \begin{tabular}[c]{@{}l@{}}achieve\\tables\\aware\\magnitude\\partitioning\\structure\\community\\complexity\\interest\\build\end{tabular}
 &

 \begin{tabular}[c]{@{}l@{}}{\color{red}algorithm}\\{\color{red}user}\\demonstrate\\{\color{red}learning}\\{\color{blue}task}\\input\\{\color{red}computing}\\{\color{red}operators}\\identify\\series \end{tabular} &
 \begin{tabular}[c]{@{}l@{}}orders\\{\color{red}aggregation}\\tools\\{\color{blue}cache}\\keyword\\{\color{blue}stored}\\{\color{blue}questions}\\{\color{blue}security}\\built\\utility\end{tabular} &

  \multicolumn{1}{r|}{\begin{tabular}[c]{@{}r@{}}data\\methods\\networks\\efficient\\representation\\set\\accuracy\\outperforms\\label\\called\end{tabular}} & \begin{tabular}[c]{@{}l@{}}due\\semantic\\stochastic\\input\\labels\\specific\\dimensional\\recognition\\strategies\\convex\end{tabular} &

  \multicolumn{1}{r|}{\begin{tabular}[c]{@{}r@{}}{\color{blue}models}\\{\color{blue}large}\\task\\{\color{blue}scale}\\solution\\level\\{\color{blue}inference}\\effectiveness\\extensive\\{\color{blue}target} \end{tabular}} & \begin{tabular}[c]{@{}l@{}}{\color{blue}matrix}\\{\color{red}latent}\\theoretical\\{\color{red}address}\\{\color{red}standard}\\compared\\{\color{red}policy}\\challenging\\prior\\{\color{red}unsupervised}\end{tabular} \\ \hline

\multicolumn{2}{|l|}{Semantic similarity}                                                                                                                                                                                                         & \multicolumn{2}{l|}{Cloud query processing}                                                                                                                                                                                                                                                    & \multicolumn{2}{c|}{Heterogeneous Relations}                                                                                                                                                                                                                                                                   & \multicolumn{2}{c|}{Game Theory}                                                                                                                                                                                                                            \\ \hline

\multicolumn{2}{|c|}{Topic 19}                                                                                                                                                                                                              & \multicolumn{2}{c|}{Topic 7}                                                                                                                                                                                                                                                             & \multicolumn{2}{c|}{Topic 40}                                                                                                                                                                                                                                                                                      & \multicolumn{2}{c|}{Topic 88}                                                                                                                                                                                                                                                                                                                    \\ \hline

\begin{tabular}[c]{@{}l@{}}search\\big\\similarity\\temporal\\objects\\scalable\\sharing\\patterns\\object\\quality\end{tabular} &
 \begin{tabular}[c]{@{}l@{}}pairs\\analyze\\communities\\step\\semantic\\log\\instance\\oriented\\nodes\\fixed\end{tabular} &

 \begin{tabular}[c]{@{}l@{}}execution\\extensive\\recent\\improve\\nodes\\cloud\\querying\\constraints\\introduce\\overhead\end{tabular} &
 \begin{tabular}[c]{@{}l@{}}interface\\enables\\variety\\requirements\\user\\fundamental\\partial\\load\\rich\\bounded\end{tabular} &

 \begin{tabular}[c]{@{}l@{}}work\\system\\constraints\\users\\local\\time\\dynamic\\markov\\key\\group\end{tabular} &
 \begin{tabular}[c]{@{}l@{}}logic\\relational\\good\\sentiment\\underlying\\resulting\\heterogeneous\\process\\active\\market\end{tabular} &

  \multicolumn{1}{r|}{\begin{tabular}[c]{@{}r@{}}problems\\approach\\search\\learning\\techniques\\applications\\general\\temporal \\previous\\games\end{tabular}} & \begin{tabular}[c]{@{}l@{}}quality\\decision\\domains\\view\\programming\\size\\reasoning\\set\\solving\\making\end{tabular} \\ \hline


   \multicolumn{4}{|c|}{ICCCN}                                                                                                                                                                                                                                                                                                                                                                                                                                                                                                        & \multicolumn{4}{c|}{ICSE}                                                                                                                                                                                                                                                                                                                                                                                                                                                                                                                                                                                                                                             \\ \hline
\multicolumn{2}{|l|}{{\color{blue}Wireless Networks}, {\color{red}Security}}                                                                                                                                                                                                         & \multicolumn{2}{l|}{{\color{blue}Cloud Computing}, {\color{red}mobility management}}                                                                                                                                                                                                                                                    & \multicolumn{2}{c|}{Programming and Code Analysis}                                                                                                                                                                                                                                                                   & \multicolumn{2}{c|}{\begin{tabular}[c]{@{}c@{}}Programming\end{tabular}}                                                                                                                                                                                                                            \\ \hline
\multicolumn{2}{|c|}{Topic 66}                                                                                                                                                                                                              & \multicolumn{2}{c|}{Topic 67}                                                                                                                                                                                                                                                             & \multicolumn{2}{c|}{Topic 13}                                                                                                                                                                                                                                                                                      & \multicolumn{2}{c|}{Topic 12}                                                                                                                                                                                                                                                                                                                    \\ \hline

\begin{tabular}[c]{@{}l@{}}{\color{blue}network}\\{\color{blue}energy} \\{\color{blue}wireless} \\{\color{blue}problem} \\scheme \\{\color{red}existing}\\{\color{red}content} \\efficient \\{\color{blue}algorithms} \\{\color{red}security}\end{tabular} &

 \begin{tabular}[c]{@{}l@{}}{\color{blue}sensor}\\{\color{red}analysis} \\{\color{red}real} \\large\\rate \\end \\state \\quality \\{\color{blue}link} \\{\color{red}mechanism}\end{tabular} &

 \begin{tabular}[c]{@{}l@{}}{\color{blue}cloud}\\{\color{blue}applications}\\design\\access\\service\\privacy \\{\color{blue}services}\\address\\low\\{\color{red}performance}\end{tabular} &
 \begin{tabular}[c]{@{}l@{}}{\color{red}level}\\{\color{blue}infrastructure}\\{\color{blue}storage}\\{\color{red}evaluation} \\experimental \\centric \\{\color{red}radio} \\{\color{blue}technique} \\{\color{red}management} \\{\color{red}approaches}\end{tabular} &

 \multicolumn{1}{r|}{\begin{tabular}[c]{@{}r@{}}code\\test\\developers\\model\\source\\system\\time\\existing\\projects\\support\end{tabular}} & \begin{tabular}[c]{@{}l@{}}cases\\language\\static\\coverage\\detection\\users\\address\\due\\evaluated\\checking\end{tabular} &

  \multicolumn{1}{r|}{\begin{tabular}[c]{@{}r@{}}program\\techniques\\real\\java\\apps\\approaches\\experiments\\art\\type\\analyze\end{tabular}} & \begin{tabular}[c]{@{}l@{}}error\\build\\reduce\\javascript\\experimental\\dependencies\\patches\\current\\typically\\github\end{tabular} \\

   \hline

\multicolumn{2}{|l|}{heterogeneous network}                                                                                                                                                                                                         & \multicolumn{2}{l|}{Communication networks}                                                                                                                                                                                                                                                    & \multicolumn{2}{c|}{Program synthesis and repair}                                                                                                                                                                                                                                                                   & \multicolumn{2}{c|}{Mining software repositories}                                                                                                                                                                                                                            \\ \hline

\multicolumn{2}{|c|}{Topic 2}                                                                                                                                                                                                              & \multicolumn{2}{c|}{Topic 56}                                                                                                                                                                                                                                                             & \multicolumn{2}{c|}{Topic 10}                                                                                                                                                                                                                                                                                      & \multicolumn{2}{c|}{Topic 20}                                                                                                                                                                                                                                                                                                                    \\ \hline

\begin{tabular}[c]{@{}l@{}}systems\\spectrum\\resources\\layer\\computing\\servers\\evaluate\\paper\\significant \\policy\end{tabular} &

 \begin{tabular}[c]{@{}l@{}}effectiveness\\case\\single\\heterogeneous\\server\\order\\support\\context \\lower\\paradigm\end{tabular} &

 \begin{tabular}[c]{@{}l@{}}propose\\communication\\framework\\simulations\\experiments\\load\\aware\\average\\demonstrate\\packets\end{tabular} &
 \begin{tabular}[c]{@{}l@{}}hop\\area\\small\\monitoring\\ratio\\amount \\characteristics\\driven\\graph\\human\end{tabular} &

 \multicolumn{1}{r|}{\begin{tabular}[c]{@{}r@{}}prediction\\issues\\effort\\traditional\\inputs\\classification\\sensitive \\theory\\community\\spreadsheets\end{tabular}} & \begin{tabular}[c]{@{}l@{}}violations\\synthesis\\capture\\automated\\needed\\end\\program\\benign\\specifications\\collaborative\end{tabular} &

  \multicolumn{1}{r|}{\begin{tabular}[c]{@{}r@{}}models\\app\\shows\\api\\search\\evolution\\participants\\errors\\popular\\mining\end{tabular}} & \begin{tabular}[c]{@{}l@{}}executions\\making\\correct\\top\\stack\\libraries\\investigate\\scalability\\questions\\thread\end{tabular} \\ \hline

\end{tabular}

}

\end{table}

%
%
%
%
%
%
%
%
%
%
%
%
%
%
%
%

\begin{figure}[htp]

  \centering

  \begin{tabular}{cc}


    \includegraphics[width=130mm]{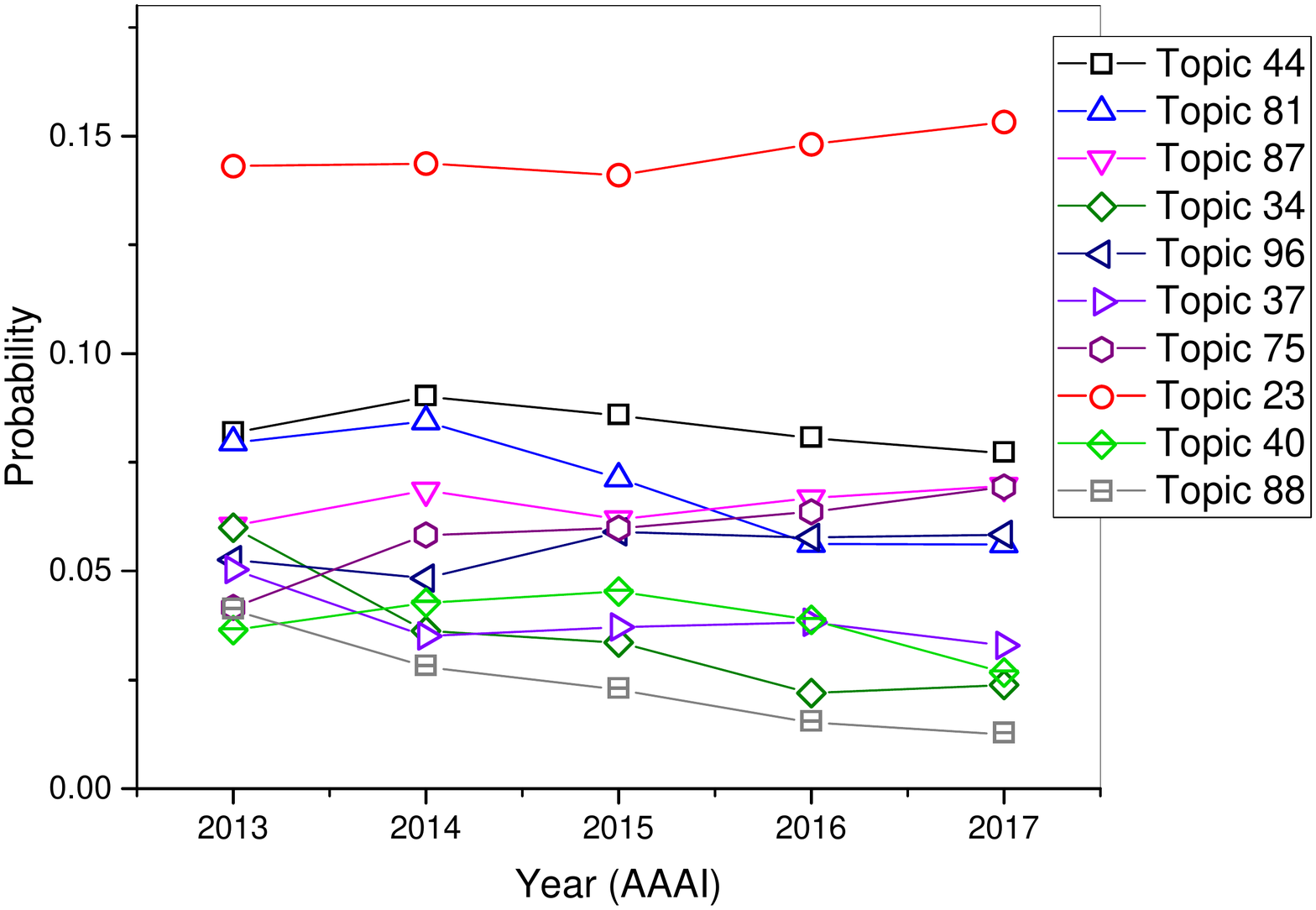}\\

    \includegraphics[width=130mm]{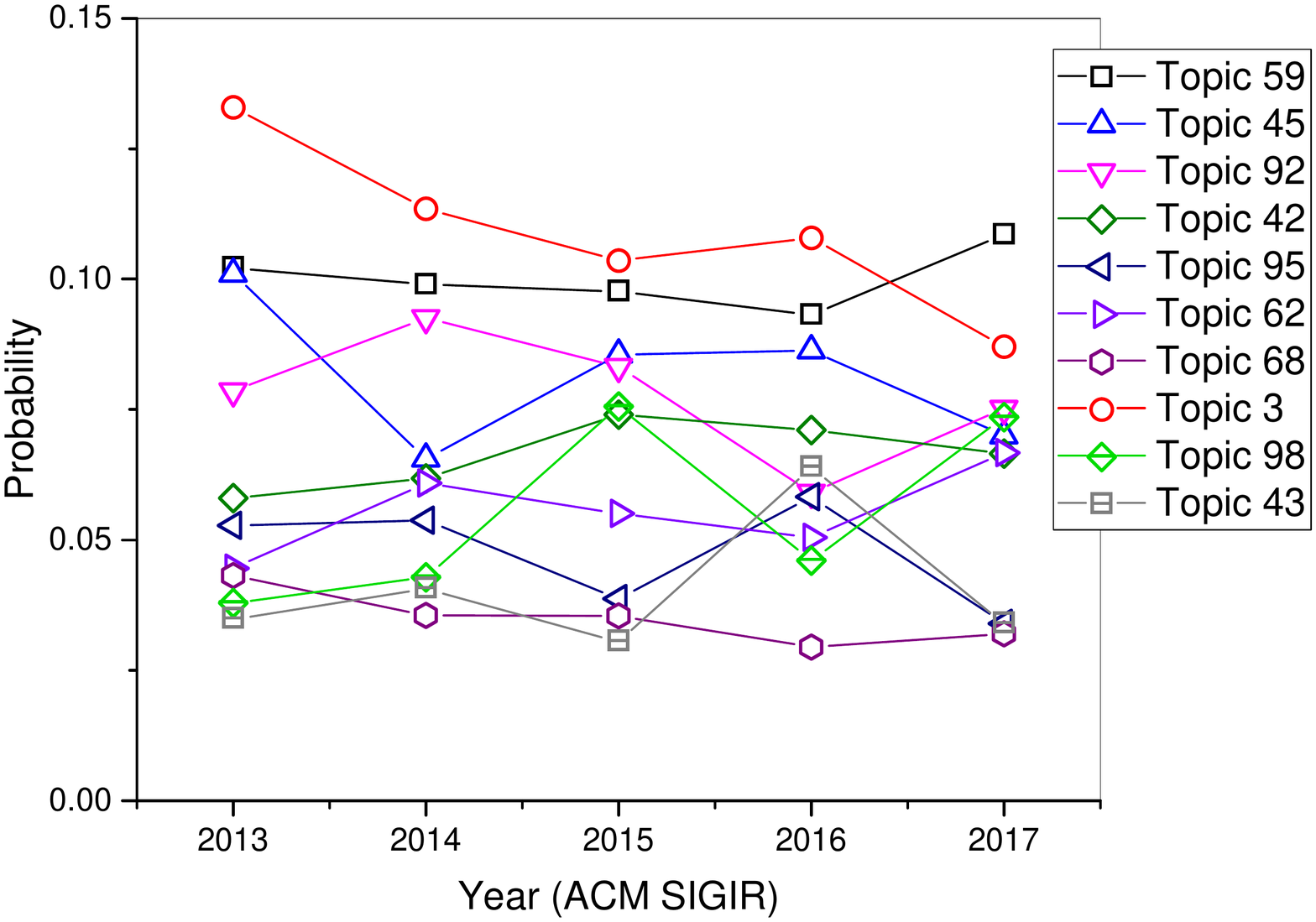}

  \end{tabular}  \protect\caption{\textbf{Temporal topic trends of AAAI and ACM SIGIR conferences, distribution of research papers in various years}}

\end{figure}

\begin{figure}[htp]

  \centering

  \begin{tabular}{cc}


   \includegraphics[width=130mm]{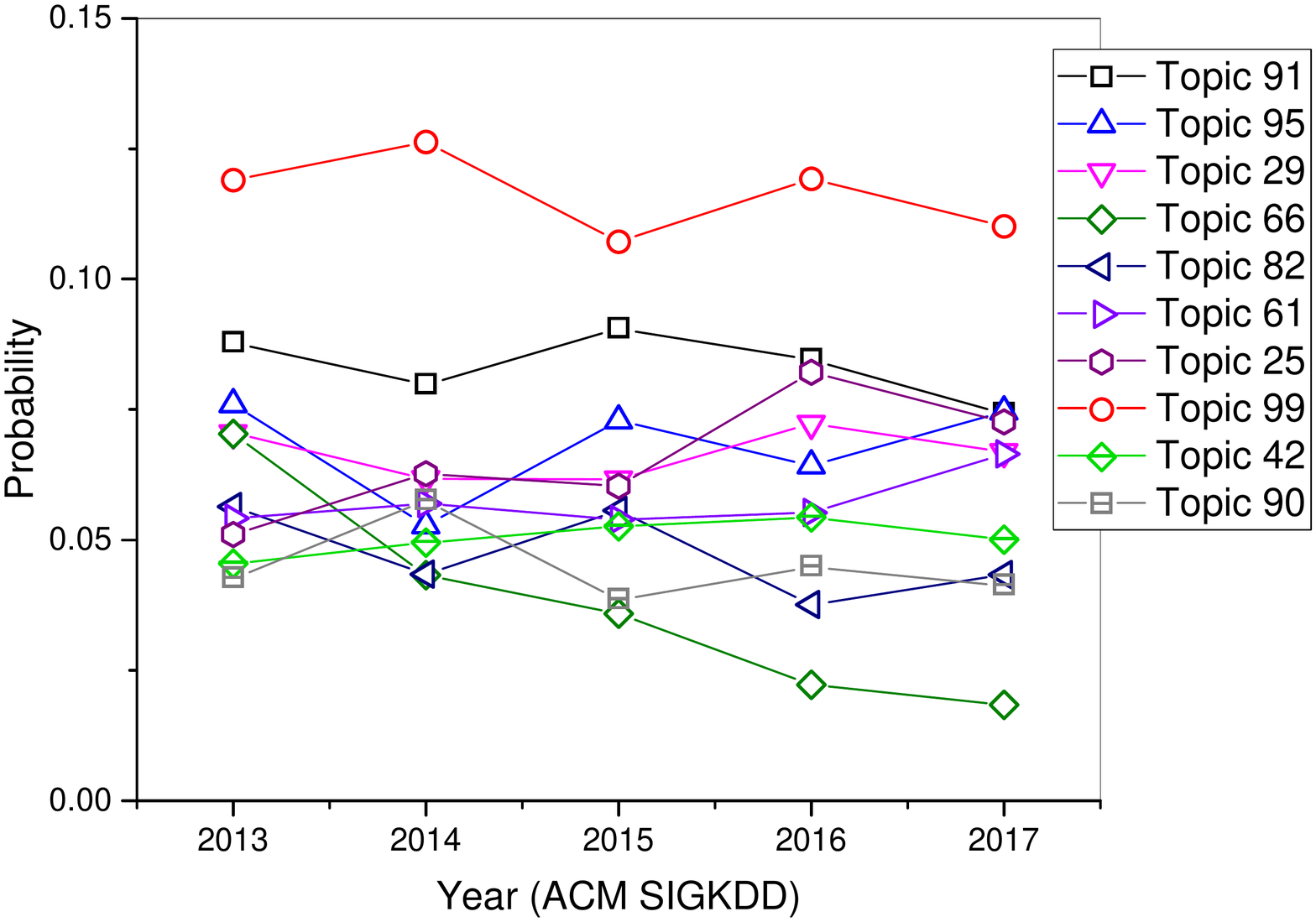}\\

   \includegraphics[width=130mm]{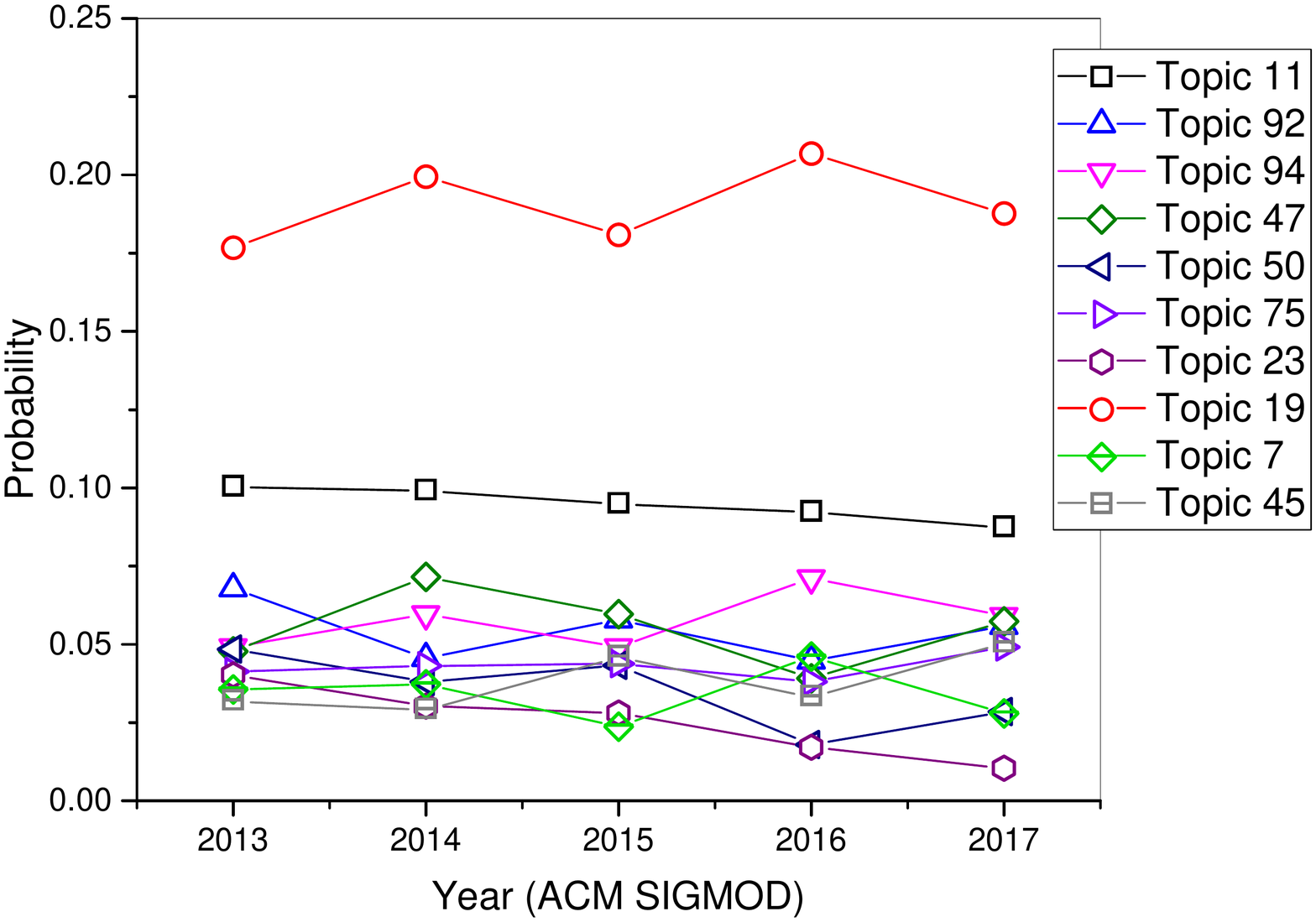}

  \end{tabular}  \protect\caption{\textbf{Temporal topic trends of ACM SIGKDD and ACM SIGMOD conferences, distribution of research papers in various years}}

\end{figure}

\begin{figure}[htp]

  \centering

  \begin{tabular}{cc}


 \includegraphics[width=130mm]{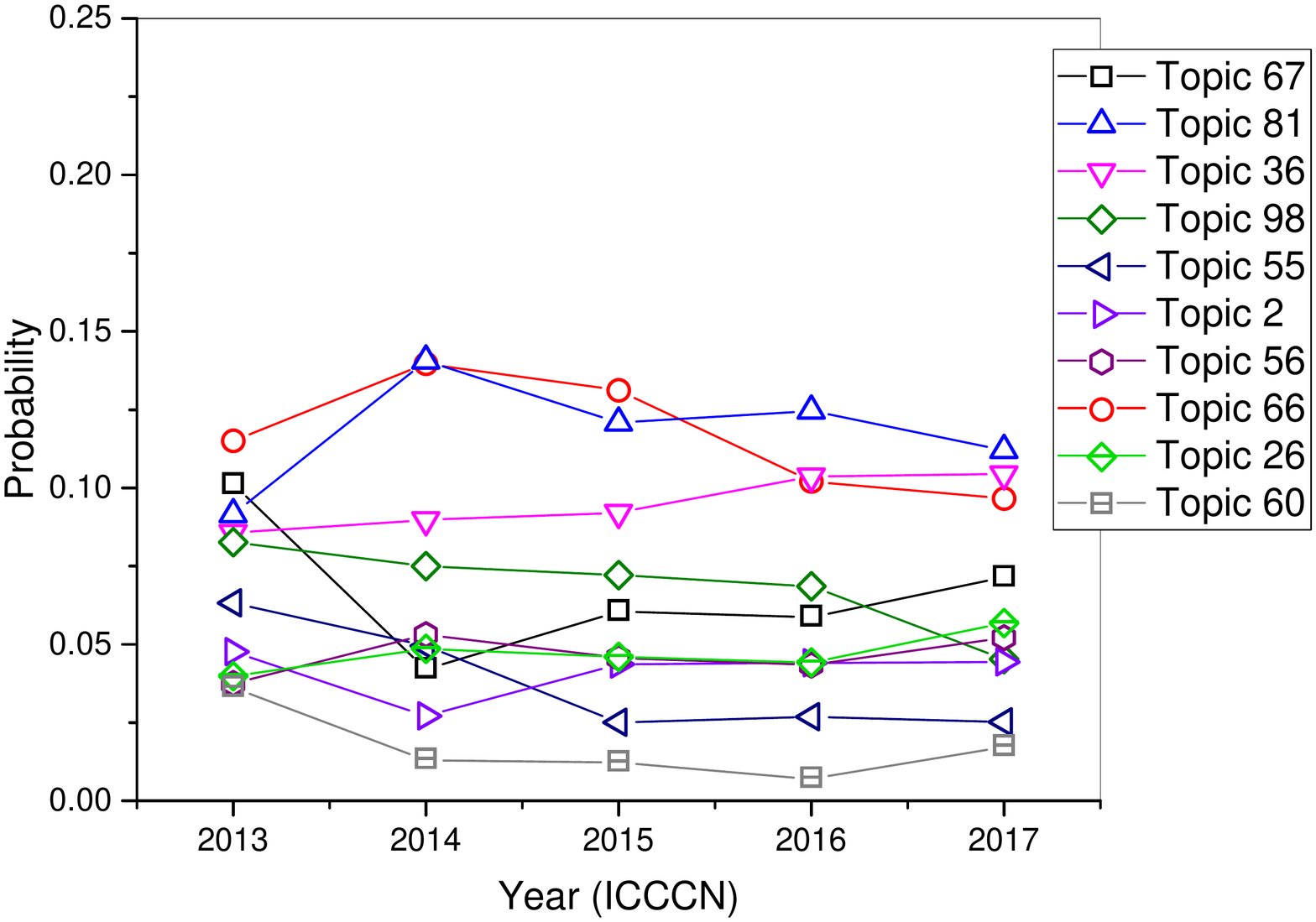}\\
   \includegraphics[width=130mm]{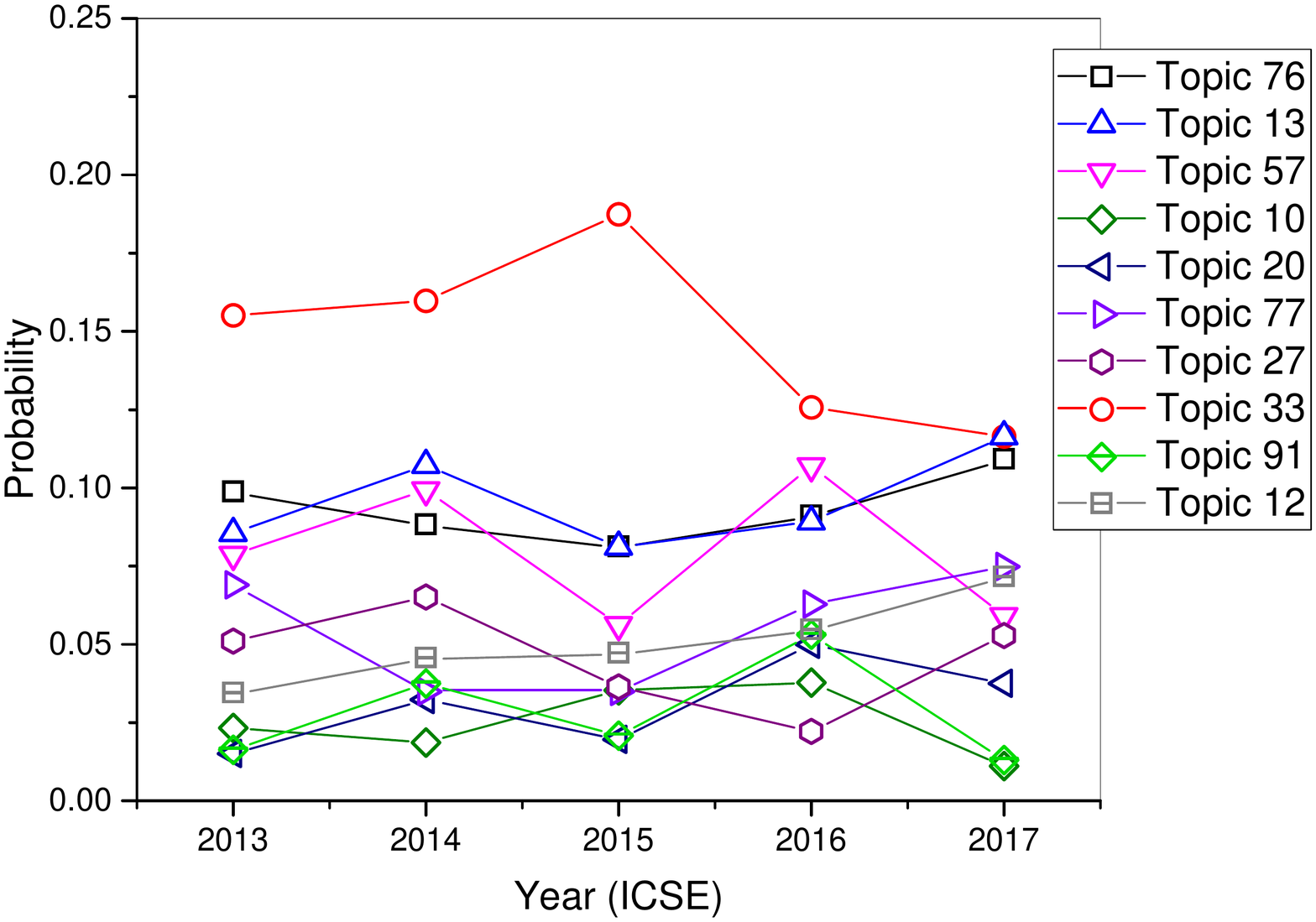}

  \end{tabular}  \protect\caption{\textbf{Temporal topic trends of ICCCN and ICSE conferences, distribution of research papers in various years}}

\end{figure}

According to Table 2 and Figure 4-6 , the following observations can be made:

In ACM SIGIR conference, Topic 59 has a trend of rising from 2016 to 2017. Topic 59 sounds considerably more generic and is consistent with 'Information Retrieval' in general, and marked by search, retrieval, queries, relevance, documents, social. Topic 62 has a high degree possibility in 2017. Moreover, the result indicated that the topic also had many interested points for researchers that focused on 'Social Media Analyses'. The derivative curves show that Topics 59, 62 and 98 in ACM SIGIR are gaining popularity in recent years.\\

In ACM SIGKDD conference, Topic 99 has a trend of rising in 2016, but falling in 2017. It indicates that the publications of this topic increase in 2016 but decreases in 2017. Topic 99 sounds considerably more generic and is consistent with 'Decision Systems and Optimization Approaches' in general, and marked by learning, systems, techniques, decision, specific, optimization. Topic 91, we can see that this topic probability curve is continuously
decreasing from 2015 to 2017, although there is a high degree possibility in 2013 that there had a lot of interested points for researchers that focused on this topic. This topic is actually a mixture of two concepts. First is 'Graph-based Algorithms', which is indicated by red words and some words are very related to each other such as model, algorithm, network, real, graph, scale. Second, it covers papers that proposed 'Approaches based on Behavior Analysis, which is indicated by blue words and some words are much related to each other such as classification, text, behavior, large, data, features. The derivative curves show that Topics 82, 61, 95 in ACM SIGKDD are gaining popularity in recent years.\\

In ACM SIGMOD, Topic 11, we can see that this topic probability curve is slowly decreasing from year 2013 to 2017. Topic 11 sounds considerably more generic and is consistent with 'Management of Data' in general, and marked by store, key,
operations, workloads, transaction, schema, cases. Topic 45, we can see that this topic has a trend of falling in 2016, but a trend of rising in 2017, and in another perspective, there is a high degree possibility in 2017 that there was a lot of interested points for researchers that focused on this topic. This topic is actually a mixture of two concepts. First is 'aggregation operators', which is indicated by red words and some words are very
related to each other such as algorithm, user, learning, operators, computing. Second, it covers papers that proposed 'Approaches based on Storage Management', which is indicated by blue words and some words are very related to each other such as aggregation, cache, stored, questions, security. The derivative curves show that Topics 50, 45, 7, 75, 92 in ACM SIGMOD are gaining popularity in recent years.\\

In AAAI conference, Topic 23, we can see that this topic probability curve is continuously increasing from 2015 to 2017. This topic sounds considerably more generic and is consistent with 'NLP and Knowledge Representation' in general, and marked by data, networks, semantic, stochastic, dimensional, and convex.

Topic 37, we can see that this topic probability curve is continuously decreasing from 2015 to 2017, although there is a high degree possibility in 2017 that there had a lot of interested points for researchers that focused on this topic. This topic is actually a mixture of two concepts. First is 'Large Scale Models', which is indicated by blue words and some words are much related to each other such as models, large, scale, inference, target, and matrix. Second, it covers papers that proposed is based on 'Policy Search and Applications of Unsupervised Learning', which is indicated by blue words and some words are very related to each other such as latent, address, standard, policy, unsupervised.

In ICCCN conference, Topic 66, this topic is actually a mixture of two concepts. First is "Approaches based on Wireless Networks", which is indicated by red words and some words are very related to each other such as network, energy, wireless, algorithms, sensor, problem, link. Second, it covers papers that proposed "Approaches based on Security in Network", which is indicated by blue words and marked by analysis, real, existing, content, security, mechanism. For Topic 67, we found this research field has a slowly increasing in curve from year 2016 to 2017, although there is a high degree possibility in 2014 that there had a lot of interested points for researchers that focused on this Topic. This topic is actually a mixture of two concepts. First is "Cloud Computing", which is indicated by blue words and some words are very related to each other such as cloud, applications, services, technique, storage, infrastructure. Second, it covers papers that proposed  'Approaches based on Mobility Management', which is indicated by red words and some words are very related to each other such as management, radio, evaluation, approaches, performance, level.\\

In ICSE conference, Topic 13, we can see that this topic probability curve is continuously increasing from 2015 to 2017. This topic sounds considerable more generic and is consistent with 'Programming and Code Analysis' in general, and marked by analysis, project, security, android, develop. Topic 12, probability curve was very smooth from year 2014 to 2016 and there was a high degree possibility in 2017 that there will be many interested points for researchers that focused on this topic. This topic covers papers that proposed algorithms based on 'Programming' and marked by error, build, program, JavaScript, patches, GitHub and Java.\\

\section{Discussion, Open Issues and Future Directions}
In this study, we analysis six conference publications from DBLP conference and used the Gibbs sampling algorithm as an evaluation parameter. We succeeded in discovering the relationship between LDA topics and paper features and also obtained the researchers' interest in research field in various years. According to our studies, some issues require further research, which can be very effective and attractive for the future.

\subsection{Behavioral and conceptual angles with two technical questions}

Technical question 1: Are all topic-words are generated from each Topic, are related to each other in each conference? Is it possible to allocate more than one concept for each topic?  According to our observation, most of topics contain some words which are common in the corpus. And other side, some words provide a simple way to find concepts, for example; word 'retrieval' can be easily recognized as a common word in the corpus about 'methods based on information retrieval', and we can find that all the topics contain this words, Topic 59 in ACM SIGIR conference (Table 2). Also some cases, Topic words may create more than one concept for each topic. For example; there are two concepts for Topic 66 in ICCCN conference. In another case, the LDA may generate irrelevant words and create just a few words for a concept. So some of the topics can be a gathering of unnecessary words, irrelevant words provide no useful information in any context and we need to be aware of and address this challenge.\\

Technical question 2: 'Ad for network' or 'Ad for advertisement', Is there an instance ambiguity or different senses for words?
One of the prominent feature of topic models is that they need no supervision in terms of data annotation. However, in some situations, limited amounts of labeled data may be available. When human read natural language text and encounter an ambiguous entity, we may make use of the context to help our understanding and choose the correct meaning between all the possibilities. For example (Topic 98, ACM SIGIR, in Table 2), the word of 'ad'! Is for 'advertisement' or 'ad hoc network'? As we can see, the word 'ad' can be related to 'advertisement' or also to 'ad hoc network'. To answer this question, it is very easy to see that topic 98 reveals 'network', 'graph', 'mobile', 'ad' and 'interactions' . If only we consider the words 'network', 'graph', 'ad' we can predict that this topic can be related to 'ad hoc network' and this topic covers papers that propose models in 'ad hoc information retrieval'. Due to this problem, a recommendation system may lead to inappropriate results.\\

\subsection{Towards development of systematic and practical recommendation systems}
Recommendation systems is increasing day-by-day with the development of scientific research, it will be more known and essential in the Internet community. Below we list a few important aspects which can be a part of future research.

\begin{itemize}

 \item Recommender systems for famous publishers, Journals and conferences are a rich source to use recommender system, and it needs more research about this field and we believe that there are many important content that we can consider as dataset, such as KDD, RecSys, IEEE fuzzy, WWW UbiComp, ACM SIGSPATIAL LBSN, AAA, ACM TIST, ACM TWEB and etc. For example, Yang and et al. proposed a novel collaborative filtering for recommendations of venues with considering extract stylometric features to obtain the similarity between papers from their writing styles and extracted 119,927 papers with considering abstract, venue information from CiteSeer Digital Library  .

\item  Hybrid approaches with machine technique, machine learning algorithms are one of the most prominent methods for text mining in computer science. Utilizing the combination of machine learning and topic modeling approaches can be effective for professional recommendation systems. As a relatively related work, we can refer to  .

\item Group recommendation, Group activities are essential components for online users in social networks. Group recommendation is a challenging problem because different group members have different preferences, and how to make a trade-off among their preferences for recommendation is still an open challenge. There are a few impressive works with considering topic modeling techniques for group recommendation that can be considered for future work  .
\end{itemize}

\section{Conclusion} \label{sec:1}
In this paper, we applied LDA algorithm and Gibbs sampling on six conference publications from the DBLP website and analyzed the behavior of researcher's interest in research field in different years. Our results showed that can discover hidden aspects to better understand the behavior of the researchers. Definitely, topic-modeling and LDA approaches can play an important role to develop the smart recommendation systems in various applications in future. Generally, recommendation systems can be an impressive interface between online users and websites in the Internet community. Certainly, the results of the article can help to scientific committee of these conferences to focus more on specific research topics for future conferences.

%
%

 \nocite{*}


\end{document}